\documentclass[a4paper]{jpconf}
\usepackage{graphicx}
\usepackage{caption}
\usepackage{pifont}
\usepackage{wrapfig}
\usepackage{hyperref}
\usepackage{amsmath}
\usepackage{cleveref}
\newcommand{\cmark}{\ding{51}}%
\newcommand{\xmark}{\ding{55}}%
\usepackage{subcaption}
\begin{document}
\title{Hierarchical Graph Neural Networks for Particle Track Reconstruction}

\author{Ryan Liu\textsuperscript{1, 2}, Paolo Calafiura\textsuperscript{1}, Steven Farrell\textsuperscript{1}, Xiangyang Ju\textsuperscript{1}, Daniel Thomas Murnane\textsuperscript{1}, Tuan Minh Pham\textsuperscript{1}}

\address{
\textsuperscript{1}Lawrence Berkeley National Laboratory,
1 Cyclotron Rd, Berkeley, CA 94720, USA\\
\textsuperscript{2}University of California,
366 Physics North, MC 7300,
Berkeley, CA 94720, USA
}

\ead{ryanliu@lbl.gov}

\begin{abstract}
We introduce a novel variant of GNN for particle tracking—called Hierarchical Graph Neural Network (HGNN).  The architecture creates a set of higher-level representations which correspond to tracks and assigns spacepoints to these tracks, allowing disconnected spacepoints to be assigned to the same track, as well as multiple tracks to share the same spacepoint. We propose a novel learnable pooling algorithm called GMPool to generate these higher-level representations called “super-nodes”, as well as a new loss function designed for tracking problems and HGNN specifically. On a standard tracking problem, we show that, compared with previous ML-based tracking algorithms, the HGNN has better tracking efficiency performance, better robustness against inefficient input graphs, and better convergence compared with traditional GNNs.
\end{abstract}

\section{Introduction}

In the upcoming High Luminosity Phase of the Large Hadron Collider (HL-LHC) \cite{Apollinari:2284929, Lyndon_Evans_2008}, the average number of inelastic proton-proton collisions per bunch $\langle\mu\rangle$ (pile-up) is expected to reach $200$ in the new silicon-only Inner Tracker (ITk). This will pose a significant challenge in track reconstruction due to the limited computational resources \cite{Collaboration:2802918}. Since charged particle reconstruction (“particle tracking”) dominates the CPU resources dedicated to event offline reconstruction, a new and efficient algorithm for event reconstruction becomes an urgent need. The HEP.TrkX project \cite{https://doi.org/10.48550/arxiv.1810.06111} and its successor the Exa.TrkX project \cite{https://doi.org/10.48550/arxiv.2003.11603} have studied Graph Neural Networks (GNNs) for charged particle tracking, and excellent performance on the TrackML dataset \cite{TrackML} has been demonstrated in Refs. \cite{Biscarat_2021, Ju_2021} and more recently on ITk simulation, referred to as GNN4ITk \cite{caillou:hal-03793565}.

However, despite the success of GNN-based tracking algorithms, there is much in these techniques that can be improved. In particular, GNN tracking suffers from two types of errors: (1) \textbf{broken tracks} (one true track split into multiple segments) and (2) \textbf{merged tracks} (a track contains spacepoints of multiple particles). In its nature, the GNN4ITk tracking pipeline prototype \cite{caillou:hal-03793565} is a process of reducing the number of edges; starting from a graph constructed for example by a multi-layer perceptron (MLP) embedding model, filter MLP and GNN edge classifiers are applied to filter out fake edges (i.e. connecting two spacepoints of distinct particles). Thus, broken tracks are more difficult to remove than merged tracks since they can only be resolved by including more edges during the graph construction stage. As such, the pipeline is very sensitive to the efficiency of the graph constructed. Furthermore, the nature of message-passing neural networks \cite{battaglia2016interaction} utilized in the GNN4ITk pipeline, precludes the passing of information between disconnected components, such as the two ends of a broken track. Broken tracks not only limit the performance of edge-cut-based algorithms but also inhibit the full capability of the message-passing mechanism.

In this paper, we present a novel machine learning model called Hierarchical Graph Neural Network (HGNN) \footnote{The code now available on \href{https://github.com/ryanliu30/HierarchicalGNN}{github}} for particle tracking to address the aforementioned problems. Similar to the pooling operation often used in Convolutional Neural Networks (CNN), the HGNN pools nodes \label{pool} into clusters called “super-nodes” to enlarge the “receptive field” of nodes to resolve the problem that a “flat” GNN cannot pass messages between disconnected components. Unlike the case of image processing where pooled pixels are already arranged on a 2D grid, the pooled super-nodes cannot use a graph induced by the original graph since disconnected components will remain disconnected. Thus we propose to utilize a K-nearest-neighbors (KNN) algorithm to build the super-graph among super-nodes to facilitate message passing between super-nodes. Furthermore, the HGNN offers us a new approach to track building, as defining a bipartite matching between nodes (spacepoints) and super-nodes (tracks). We measure the performance of this matching procedure against several baselines and show that it can not only recover broken tracks, but also produces fewer fakes tracks from merging.

\begin{figure}
    \hfill
    \begin{subfigure}[b]{0.3\textwidth}
         \centering
         \includegraphics[width=\textwidth]{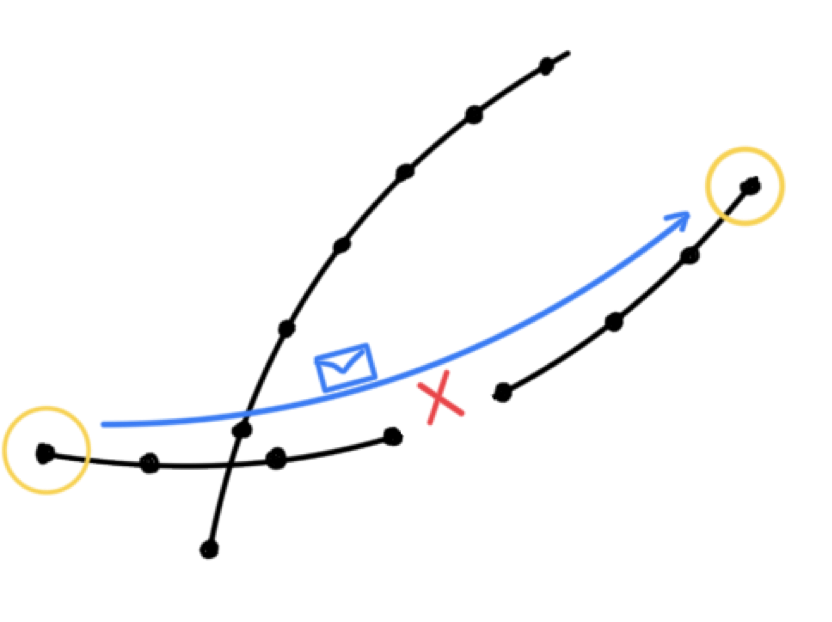}
         \caption{Flat GNN}
         \label{fig:flat GNN}
    \end{subfigure}
    \hfill
    \begin{subfigure}[b]{0.35\textwidth}
         \centering
         \includegraphics[width=\textwidth]{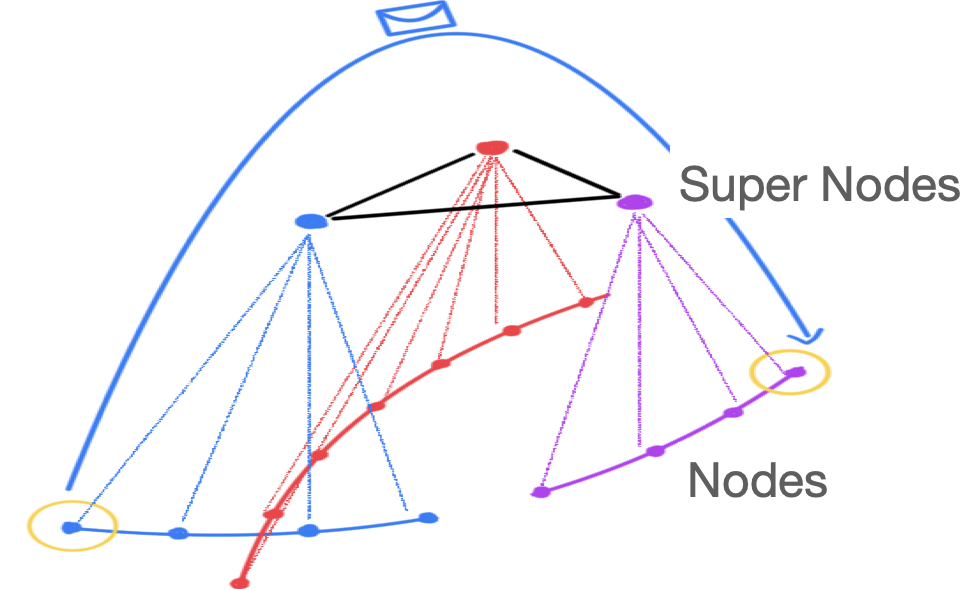}
         \caption{Hierarchical GNN}
         \label{fig:HGNN}
    \end{subfigure}
    \hfill
    \hfill
    \caption{The HGNN can not only shorten the distance between two nodes and effectively enlarge the receptive field but also pass messages between disconnected components}
    \label{fig:flat vs HGNN}
\end{figure}

\section{Related Work}
\subsection{The GNN4ITk Pipeline for Charged Particle Tracking}
The GNN4ITk pipeline \cite{Ju_2021, caillou:hal-03793565} aims to accelerate particle tracking by utilizing geometric deep learning models. The pipeline as implemented can be divided into four steps: firstly, graph construction takes place to build a graph on the input point-cloud. With one possible construction technique, an MLP is trained to embed spacepoints into a high-dimensional space such that spacepoints belonging to the same particle gets closer in space; a fixed radius graph is then built and passed to a “filter” MLP. The filter takes in spacepoint doublets and prunes the graph down by a $O(10)$ factor in the number of edges. A graph neural network is used to prune the graph further down. Finally, the tracks are built by running a connected components algorithm on the pruned graphs, and ambiguities are resolved by a walk-through algorithm based on topological sorting.
 
\subsection{Graph Pooling Algorithms}
As discussed in \cref{pool}, the pooling algorithm is a crucial piece of the HGNN architecture. Graph pooling has long been studied in the context of graph neural networks as generating graph representations require some global pooling operation. Ying \textit{et al.} introduced DiffPool \cite{https://doi.org/10.48550/arxiv.1806.08804}, which pools the graph by aggregating nodes according to weights generated by a GNN. DiffPool pools the graph to a fixed number of super-nodes, and the pooled graph has a dense adjacency matrix. Lee \textit{et al.} proposed SAGPool \cite{lee2019self}, which pools a graph by selecting top-k rank nodes and uses the subgraph induced. However, SAGPool does not support soft assignment, i.e. assigning a node to multiple super-nodes. The granularity is completely defined by the hyperparameter $k$ and thus also pools to a fixed number of super-nodes. Diehl proposed EdgePool \cite{diehl2019edge}, which greedily merges nodes according to edge scores. It is capable of generating a graph that is sparse and variable in size. These pooling algorithms and their features are presented in \cref{tab:compare}, along with our proposed pooling technique, described in \cref{sec:GMPool}.
\begin{table}[ht]
\caption{Graph Pooling Algorithms}
\centering
\begin{tabular}{p{3.5cm}p{3cm}cccc}
\br
Tracking Goal & Feature & DiffPool & SAGPool & EdgePool & GMPool (ours) \\
\mr
Subquadratic scaling & Sparse & \xmark & \cmark & \cmark & \cmark\\
End-to-end trainable & Differentiable & \cmark & \cmark & \cmark & \cmark\\
Variable event size & Adaptive number of clusters & \xmark & \xmark & \cmark & \cmark\\
Many hits to many particles relationship & Soft assignment & \cmark & \xmark & \xmark & \cmark\\
\br
\end{tabular}\label{tab:compare}
\end{table}

\subsection{Hierarchical Graph Neural Networks}
Hierarchical structures of graph neural networks have been studied in the context of many graph learning problems; some of them utilize deterministic pooling algorithms or take advantage of preexisting structures to efficiently create the hierarchy \cite{Zhang_Zhuang_Zhu_Shi_Xiong_He_2020, 9101846, guille2022document, 10.1093/nar/gkab044, zhong2022hierarchical}, while the others \cite{gao2019graph, rampavsek2021hierarchical, chen2021hierarchical} create the hierarchy in a learnable fashion. Compared with solely graph pooling operations \cite{https://doi.org/10.48550/arxiv.1806.08804}, by retaining both pooled and original representations one has the capability of simultaneously performing node predictions and learning cluster-level information. Furthermore, as shown in \cite{rampavsek2021hierarchical}, introducing hierarchical structures can solve the long-existing problem of the incapability of capturing long-range interactions in graphs. Empirical results also show that Hierarchical GNNs have better convergence and training stability compared with traditional flat GNNs.
\section{Model Architecture}
In order to build the model, there are several challenges that must be tackled, namely, pooling the graph, message passing in the hierarchical graph, and designing a loss function for such a model. In the following section, we introduce our proposed methods for each of them.
\begin{figure}
    \centering
    \includegraphics[width=0.9\textwidth]{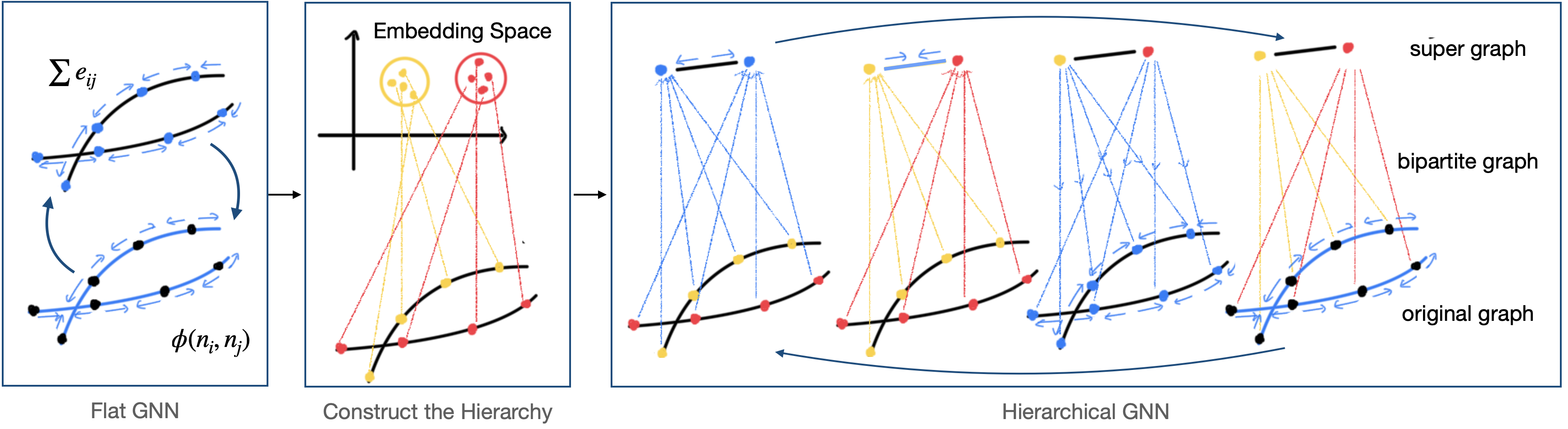}
    \caption{A schematic overview of the HGNN architecture. A flat GNN encoder is used to transform features and embed spacepoints. A pooling algorithm (GMPool) follows to build the hierarchy using the embedded vectors. Finally, hierarchical message passing is applied iteratively to obtain final representations of both nodes and super-nodes.}
    \label{fig:my_label}
\end{figure}
\subsection{Gaussian Mixture Pooling}\label{sec:GMPool}
\begin{figure}
    \hfill
    \begin{subfigure}[b]{0.65\textwidth}
         \centering
         \includegraphics[width=\textwidth]{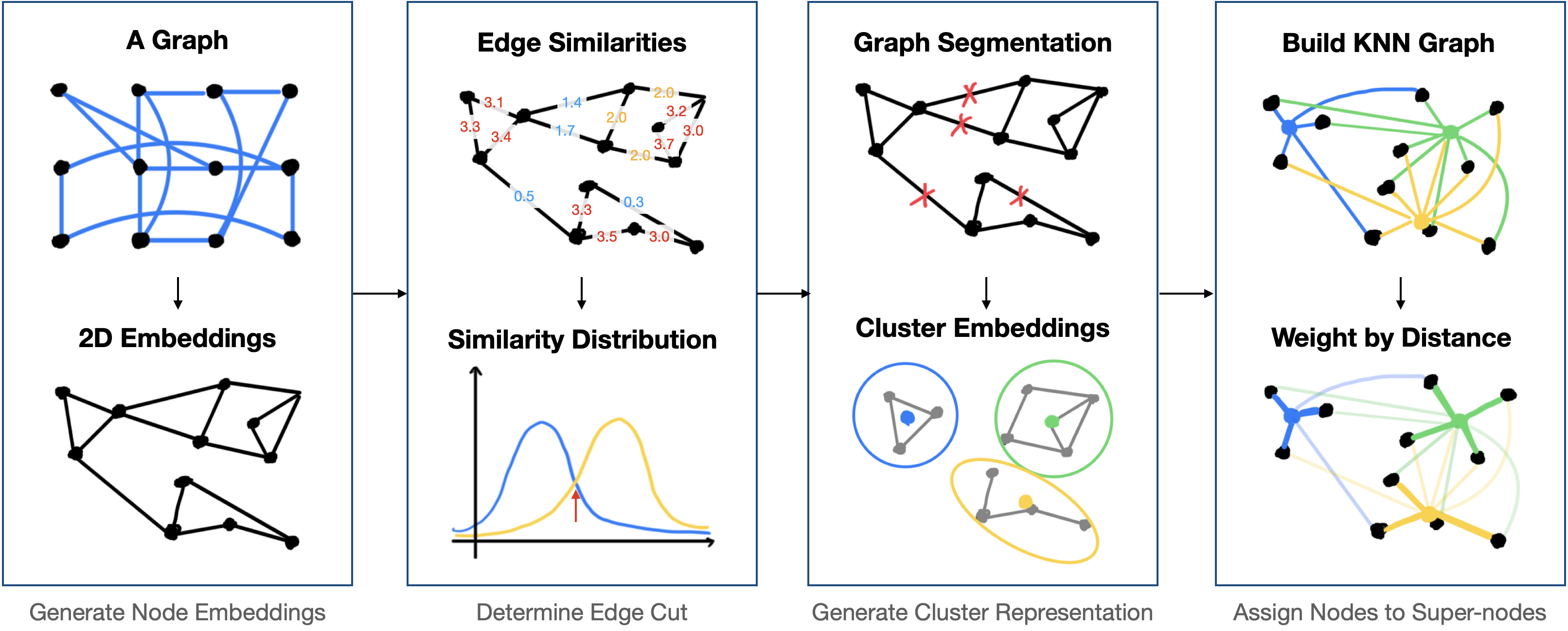}
         \caption{}
         \label{fig:GMPool-overview}
    \end{subfigure}
    \hfill
    \begin{subfigure}[b]{0.28\textwidth}
         \centering
         \includegraphics[width=\textwidth]{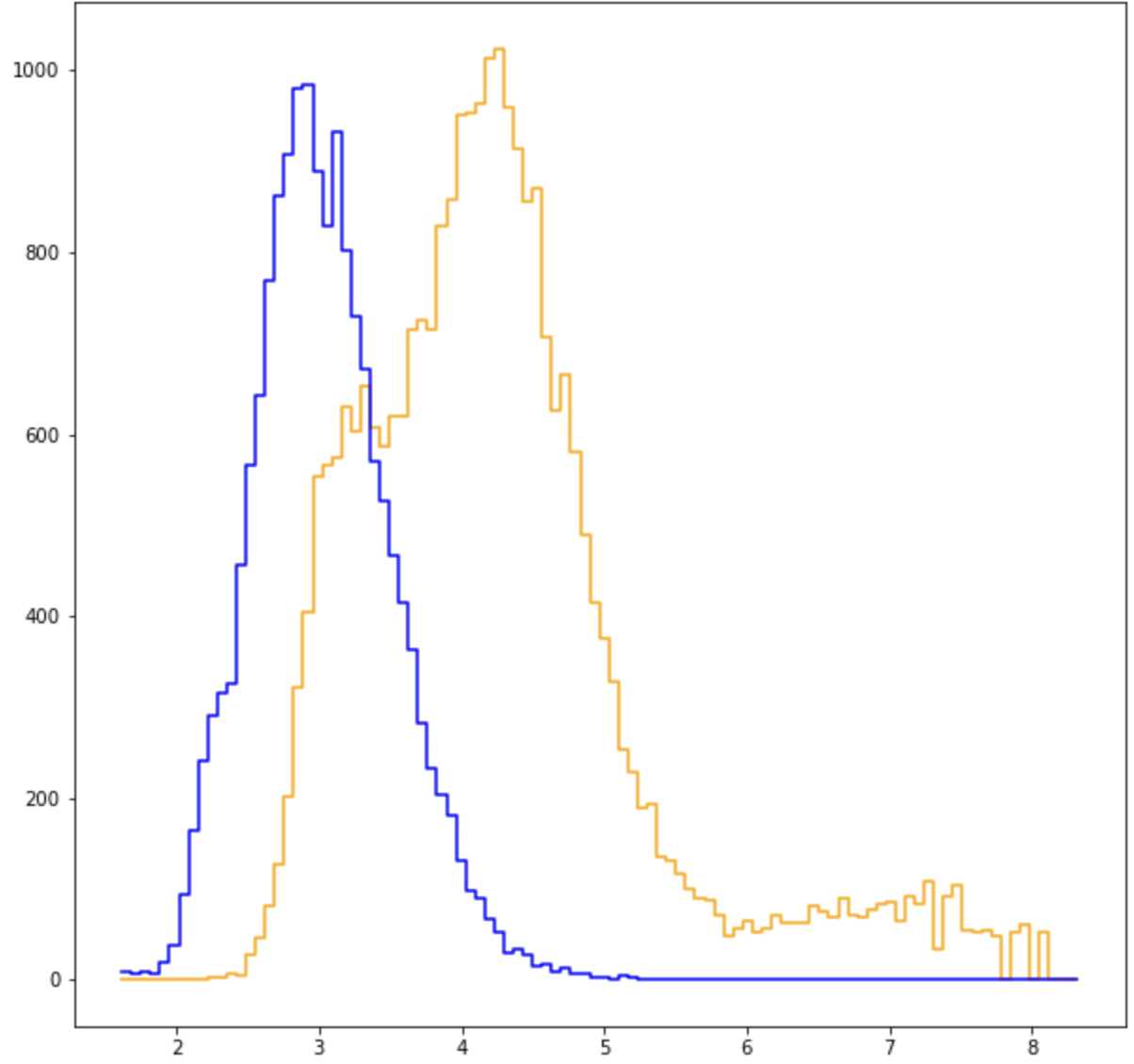}
         \caption{}
         \label{fig:sdist}
    \end{subfigure}
    \hfill
    \caption{(a): schematic overview of the GMPool algorithm. (b): Distribution of edge similarities. Edges connecting spacepoints of the same particle are colored in yellow and otherwise blue.}
    \label{fig:GMPool}
\end{figure}

In order to provide the features in \cref{tab:compare}, we propose a method that leverages the connected components algorithm and Gaussian Mixture Model. The algorithm takes a set of node embeddings as input. The embeddings are then used to calculate edge-wise similarities defined as $s_{ij}=\tanh^{-1}(\vec v_i\cdot \vec v_j)$. We hypothesize that the graph consists of two types of edges, in-cluster edges and out-of-cluster edges. Then, given the distribution of node similarities, we fit a Gaussian Mixture Model (GMM) to obtain the estimation of the in-cluster and out-of-cluster distributions $p_{in}(s)$ and $p_{out}(s)$. An example distribution is plotted in \cref{fig:sdist}. We then solve for $s_{cut}$ by $\ln(p_{in}(s_{cut})) - \ln(p_{out}(s_{cut})) = r$, where $r$ is a hyperparameter defining the resolution of the pooling algorithm. The $s_{cut}$ value that gives the best separation of in- and out-of-cluster Gaussians is chosen, and edges with scores below this value are cut. The connected components algorithm follows, and the components $C_\alpha$ of the cut graph are regarded as super-nodes. 

To construct super-edges, first super-node embeddings are defined as the centroid of each of the connected components in the embedding space, i.e. $\vec V_\alpha = \frac{\vec V'_\alpha}{\left\Vert\vec V'_\alpha\right\Vert_{L_2}}$ where $\vec V'_\alpha=\frac{1}{N(C_{\alpha})}\sum_{i\in C_{\alpha}}\vec v_i$
To connect nodes with super-nodes, similar to the method used in \cite{qasim2019learning}, we maintain the sparsity by constructing the bipartite graph with the k-nearest neighbors algorithm. The differentiability can be restored by weighting each of the edges according to the distance in the embedding space, i.e. and $w_{i\alpha}=\frac{\exp(v_i\cdot V_\alpha)}{\sum_{\alpha\in \mathcal{N}(i)}\exp(v_i\cdot V_\alpha)}$
Finally, node features are aggregated to be super-node features according to the graph weights.
The super-graph construction is identical except that the k-nearest neighbor search has the same source and destination set. Thanks to its edge-cut nature, the GMPool has sub-quadratic time complexity and runs in milliseconds on our graphs.
\subsection{Hierarchical Message Passing Mechanism}
In general, it is possible to stack arbitrarily many pooling layers to obtain a hierarchy of arbitrary height. However, the nature of tracking problems suggests that a spacepoint-particle hierarchy will be sufficient for tracking problems. Thus, the pooling layer in this work is kept to be of two levels. For each of the nodes, we update it by aggregating adjacent edge features, super-nodes features weighted by bipartite graph weights, and its own features. For each of the super-nodes, it is updated by aggregating super-edge features weighted by super graph weights, node features weighted by bipartite graph weights, and its own features. For edges and super-edges, their update rule is identical to the one used in interaction networks.
\subsection{Bipartite Classification Loss}
At this point, the architecture of HGNN is possible to train on traditional tasks such as node-embedding thanks to GMPool's differentiability. This feature is useful for apples-to-apples comparisons between flat and hierarchical GNNs under the same training regimes. However, to exploit the full potential of the HGNN, we propose a new training regime for it specifically. The most natural way of doing track labeling with HGNN is to use super-nodes as track candidates. For each of the spacepoint-track pairs (bipartite edges), a score is produced to determine if it belongs to a specific track. A maximum-weight bipartite matching algorithm is used to match tracks to super-nodes to define the “truth” for each of the bipartite edges. The loss is given by the binary cross-entropy loss defined by the matched truth. An auxiliary hinge embedding loss is also used for the first warm-up epochs to help the embedding space stably initialize. 
\section{Results}
\subsection{Dataset}
In this paper, the dataset used to report the performance of HGNN is that of the TrackML Challenge\cite{TrackML}. The TrackML dataset contains events of simulated proton-proton collisions at $\sqrt{s} = 14\mathrm{TeV}$ with pile-up $\langle\mu\rangle = 200$. Details can be found in \cite{TrackML}. The HGNN has been evaluated in two scenarios; the first scenario is called TrackML-full and contains 2200 filter-processed events, each with approximately $O(7k)$ particles and $O(120k)$ spacepoints. In addition to that, an extensive test of robustness has been done on Bipartite Classifiers, using a simplified dataset TrackML-1GeV. We take the subgraph induced by removing any track below $p_T=1\mathrm{GeV}$. Such an event typically consists of $O(1k)$ particles and $O(10k)$ spacepoints.  
\subsection{Evaluation}
The evaluation metric is tracking efficiency and purity. A particle is matched to a track candidate if \textbf{(1)}: the track candidate contains more than 50\% of the spacepoints left by the particle and \textbf{(2)}: more than 50\% of the spacepoints in the track candidate are left by the particle. A track is called reconstructable if it \textbf{(1)} left more than 5 spacepoints in the detector and \textbf{(2)} has $p_T\geq 1\mathrm{GeV}$. The tracking efficiency and fake rate (FR) are thus defined as:
\begin{align*}
    \mbox{Eff} := \frac{N(\mbox{matched}, \mbox{reconstructable})}{N(\mbox{reconstructable})} && \mbox{FR} := 1 - \frac{N(\mbox{matched})}{N(\mbox{track candidates})}
\end{align*}

\subsection{Experiments}
We evaluate four models on the TrackML-full dataset. \textbf{(1)}: Embedding Flat GNN (E-GNN), \textbf{(2)}: Embedding Hierarchical GNN (E-HGNN), \textbf{(3)}: Bipartite Classifier Hierarchical GNN (BC-HGNN), \textbf{(4)}: Edge Classifier Flat GNN (EC-GNN). The first two serve for apples-to-apples comparisons between flat and hierarchical GNNs - the loss function is the same as the hinge embedding loss used for the metric learning graph construction; tracks candidates are selected by applying a spatial clustering algorithm (H-DBSCAN). The third model represents the state-of-the-art hierarchical GNN for particle tracking; the last one is identical to the GNN4ITk pipeline, and serves as a baseline. The performance of a truth-level connected-components (Truth-CC) track builder are also reported; this takes in filter-processed graphs and prunes them down with ground truth. It is a measure of the graph quality and also an upper bound of edge classifier flat GNN performance. The timing results are obtained on a single Nvidia A100 GPU. To test robustness against edge inefficiency, we remove 0\%, 20\%, 30\%, and 40\% of the edges and train the Bipartite Classifier model to compare it with the Truth-CC. 
\begin{table}[h]
\caption{TrackML-Full experiment results. Comparison between embedding models shows that hierarchical structure can enhance the expressiveness of GNNs. Comparing Bipartite Classifiers with the Truth CC, we can see that Bipartite Classifiers can recover some of the tracks that cannot be reconstructed by edge-based GNNs\protect\footnotemark. The timing results also show that HGNN scales to large input graphs of HL-LHC events competitively with other embedding GNNs}
\centering
\begin{tabular}{llllll}
\br
Models & E-GNN & E-HGNN & BC-HGNN & EC-GNN & Truth-CC\\
\mr
Efficiency & 94.61\% & 95.60\% &\textbf{97.86\%} & 96.35\% & 97.75\% \\
Fake Rate & 47.31\% & 47.45\% & \textbf{36.71\%} & 55.58 \% & 57.67\% \\ 
Time (sec.) & 2.17 & 2.64 & 1.07 & \textbf{0.22} & 0.07\\
\br
\end{tabular}
\end{table}
\begin{table}[h]
\caption{TrackML-1GeV extensive robustness test results. We can see that Bipartite Classifiers (BC) are very robust against inefficiencies, whereas edge-based GNN's performance is strongly influenced by missing edges.}
\centering
\begin{tabular}{lllllll}
\br
Percent Edge Removed & 0\% &10\% & 20\% & 30\% & 40\% & 50\%\\
\mr
BC Efficiency & 98.55\% & 98.39\% & 97.68\% & 96.63\% & 95.10\% & 92.79\%\\
BC Fake Rate & \phantom{0}1.23\% & \phantom{0}1.55\% & \phantom{0}2.13\% & \phantom{0}3.10\% & \phantom{0}4.75\% & \phantom{0}7.31\%\\
Truth-CC Efficiency & 98.72\% & 96.21\% & 92.31\% & 85.81\% & 77.26\% & 64.81\%\\
Truth-CC Fake Rate & \phantom{0}5.87\% & 15.53\% & 24.40\% & 33.48\% & 42.99\% & 53.12\%\\
\br
\end{tabular}
\end{table}
\footnotetext{The fake rates given here are not intended as final track building performance, but rather as a model-to-model comparison. A walkthrough algorithm such as in the GNN4ITk pipeline, as well as track cleaning and ambiguity resolution can greatly reduce final fake rates.}
\section{Conclusion}
In this paper, we introduced a novel graph neural network called a hierarchical graph neural network. We also proposed a new learnable pooling algorithm called GMPool to construct the hierarchy. The architecture successfully resolved the issues of GNN being incapable of capturing long-range interactions and the GNN particle tracking pipeline being sensitive to graphs' efficiency. Creating higher-level representations both shortens the distance between distant nodes in graphs and offers new methods of building track candidates. Empirical results demonstrate that Hierarchical GNNs have superior performance compared with flat GNNs. The hierarchical GNN is available at \href{repo}{https://github.com/ryanliu30/HierarchicalGNN} and has been integrated into the common framework of the GNN4ITk pipeline \cite{GNN4ITk}.
\section{Acknowledgements}
This research was supported in part by: the U.S. Department of Energy’s Office of Science, Office of High Energy Physics, under Contracts No. DE-AC02-05CH11231 (CompHEP Exa.TrkX). This research used resources of the National Energy Research Scientific Computing Center (NERSC), a U.S. Department of Energy Office of Science User Facility located at Lawrence Berkeley National Laboratory, operated under Contract No. DE-AC02-05CH11231.
\section*{References}
\bibliographystyle{iopart-num}
\bibliography{refs}
\end{document}